\documentclass[aps, pra,twocolumn,showpacs,amsmath,amssymb,preprintnumbers,superscriptaddress,10pt]{revtex4-2}
\usepackage{appendix}
\usepackage{booktabs}
\usepackage{tikz}
\usetikzlibrary{shapes.geometric,shapes.arrows,decorations.pathmorphing}
\usetikzlibrary{matrix,chains,scopes,positioning,arrows,fit}
\usepackage{circuitikz}
\usepackage{pgfplots}
\pgfplotsset{compat=1.15}
\usepgfplotslibrary{units, groupplots, external, fillbetween}
\usepackage{graphicx}
\usepackage{braket}
\graphicspath{{figures/}}
\usepackage{siunitx}
\usepackage{hyphenat}
\usepackage[colorlinks,bookmarks=false]{hyperref}
\usepackage{color}

\usepackage[capitalise]{cleveref}
\crefname{section}{Sec.}{Secs.}
\Crefname{section}{Section}{Sections}

\definecolor{pink}{RGB}{255,0,255}

\definecolor{fakedRed}{rgb}{0,0,1}

\begin{document}
	
\title{Strong pulse illumination hacks self-differencing avalanche photodiode detectors in a high-speed quantum key distribution system}

\author{Binwu~Gao}
\affiliation{Institute for Quantum Information \& State Key Laboratory of High Performance Computing, College of Computer Science and Technology, National University of Defense Technology, Changsha 410073, China}
	
\author{Zhihao~Wu}
\affiliation{Institute for Quantum Information \& State Key Laboratory of High Performance Computing, College of Computer Science and Technology, National University of Defense Technology, Changsha 410073, China}
	
\author{Weixu~Shi}
\affiliation{Institute for Quantum Information \& State Key Laboratory of High Performance Computing, College of Computer Science and Technology, National University of Defense Technology, Changsha 410073, China}
	
\author{Yingwen~Liu}
\affiliation{Institute for Quantum Information \& State Key Laboratory of High Performance Computing, College of Computer Science and Technology, National University of Defense Technology, Changsha 410073, China}
	
\author{Dongyang~Wang}
\affiliation{Institute for Quantum Information \& State Key Laboratory of High Performance Computing, College of Computer Science and Technology, National University of Defense Technology, Changsha 410073, China}
	
\author{Chunlin~Yu}
\affiliation{China Greatwall Research Institute, China Greatwall Technology Group CO.,\  LTD.\ 518057, shenzhen, China}
	
\author{Anqi~Huang}
\email[]{angelhuang.hn@gmail.com}
\affiliation{Institute for Quantum Information \& State Key Laboratory of High Performance Computing, College of Computer Science and Technology, National University of Defense Technology, Changsha 410073, China}
	
\author{Junjie~Wu}
\email[]{junjiewu@nudt.edu.cn}
\affiliation{Institute for Quantum Information \& State Key Laboratory of High Performance Computing, College of Computer Science and Technology, National University of Defense Technology, Changsha 410073, China}

\date{\today}
	
	\begin{abstract}
		Implementation of high-speed quantum key distribution~(QKD) has become one of the major focuses in the field, which produces high key-generation rate for applications. To achieve high-speed QKD, tailored techniques are developed and employed to quickly generate and detect quantum states. However, these techniques may introduce unique loopholes to compromise the security of QKD systems. In this paper, we investigate the loopholes of self-differencing~(SD) avalanche photodiode~(APD) detector, typically used for high-speed detection in a QKD system, and demonstrate experimental testing of SD APD detector under strong pulse illumination attack. This attack presents blinding stability and helps an eavesdropper to learn the secret key without introducing extra QBER. Based on this testing, we propose a set of criteria for protecting SD APD detectors from the strong pulse illumination attack.
	\end{abstract}
	\maketitle
	
	\section{INTRODUCTION}
		Quantum key distribution~(QKD), whose security is guaranteed by the laws of quantum mechanics, allows two remote and legitimate users to share a private and secret key~\cite{bennett1984,ekert1991,bennett1992a,lo2012}. Nowadays, for the need of high-speed key generation rate, prepare-and-measure QKD protocols~\cite{bennett1984,inoue2002,scarani2004,stucki2005} are the common choice, instead of measurement-device-independent~(MDI) QKD protocol~\cite{lo2012} that removes all security loopholes in measurement devices but has relative low rate. In order to achieve high key generation rate with help of system's high repetition frequency, traditional gated avalanche photodiode~(APD) detector might not be suitable due to the effect of afterpulse noise that is produced by trapped avalanche charge. To reduce the afterpulse noise, it is required that the weaker avalanche signal shall be sensed, which can be satisfied by employing self-differencing~(SD) technique to a APD. Therefore, the SD APD detector is commonly used in gigahertz high-speed QKD systems~\cite{yuan2008,Dixon2008}. 
		
		Although QKD has been proved to be information-theoretically secure in theory, there are still some loopholes in practical implementation~\cite{brassard2000,lydersen2010a,xu2010,gerhardt2011,bugge2014,huang2016,Sajeed2016,huang2019a,Chistiakov2019,huang2020}. For example, the single-photon detectors(SPDs), which are the core devices in BB84 QKD systems, may be hacked by the eavesdropper Eve via the after-gate attack~\cite{wiechers2011}, the time-shift attack~\cite{qi2007,zhao2008}, the detector-blinding attack~\cite{lydersen2010a,gerhardt2011}, and so on. In order to defend against these attacks on the detection devices, security patches~\cite{yuan2011,silva2012,lim2015} are effective countermeasures. That is, once a new type of attack is discovered, a corresponding countermeasure against this attack may be proposed and realized in an existing QKD system~\cite{Xu2020}. Recently, in order to ensure the most secure conditions to operate SD APD detectors in QKD systems, a set of so called ``best-practice criteria'' for practical security of SD APD detectors has been proposed~\cite{koehlersidki2018}. 
		
		Continuous-wave~(c.w.) light is usually regarded to achieve a reliable eavesdropping, so as for the ``best-practice criteria'' that only considers the case of c.w. blinding attacks~\cite{koehlersidki2018}. Instead, the power fluctuation of optical pulse may expose the hacking behaviour of an eavesdropper~\cite{koehlersidki2018,jiang2013}. However, in this study, we find that strong pulse illumination attack presents blinding stability. By using strong optical pulse, eavesdropper can blind the SD APD detector continuously and steadily without introducing extra quantum bit rate error~(QBER). 
		
		In this paper, under the practice criteria~\cite{koehlersidki2018}, we experimentally demonstrate that the SD APD detector in a QKD system can be directly blinded by using strong optical pulses with the repetition frequency as the same as the gating frequency of the SD APD detector. Then we trigger SD APD detector when it is completely blinded and realize the control in the detection probability of the detector from 0\% to 100\%. This study shows that the SD APD detector can be successfully hacked by the pulse illumination attack, which might compromise the security of a high-speed QKD system with SD APD detectors. Afterward, we propose a set of criteria for practical security of SD APD detectors by taking the threat of pulse illumination attack into accounts.
		
		The paper is structured as follows. Section~\ref{Ⅱ} introduces the operation principle of SD APD detectors and the general process of strong pulse illumination attack. The experimental setup and selection criteria of discrimination level of the tested SD APD detector are described in Sec.~\ref{Ⅲ}. Under the practice criteria, the methodology and testing results of pulse illumination attack are presented in Sec.~\ref{Ⅳ}. In Sec.~\ref{Ⅴ}, we show the difference between the pulse illumination attack in this work and the previous attack on SD APD detector disclosed in Ref.~\cite{jiang2013}, analyze the incomprehensiveness of practical criteria in Ref.~\cite{koehlersidki2018} and propose a list of practical criteria to resist pulse illumination attack. Finally, we conclude in Sec.~\ref{VI}.
	
	\section{WORKING PRINCIPLE OF SD APD DETECTORS AND STRONG PULSE ILLUMINATION ATTACK} \label{Ⅱ}
		In this part, we firstly introduce the operation principle of SD APD detectors by taking the SD APD detector tested in this study as an example. Then we introduce general process of strong pulse illumination attack that proposed in this work.
	
		\begin{figure}[htbp]
			\centering
			\includegraphics[width=\linewidth]{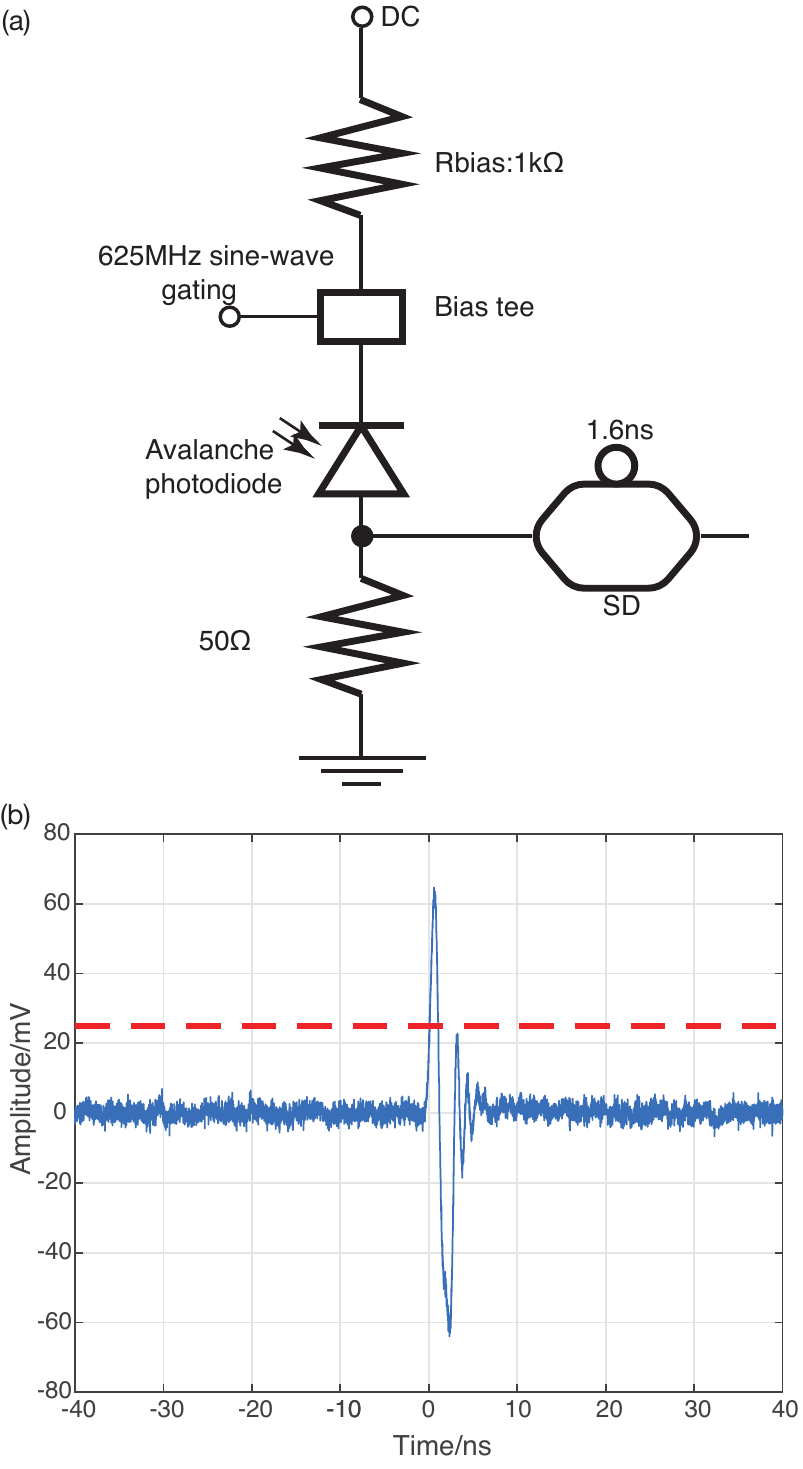}%
			\caption{(a) The schematic circuit of the tested SD APD detector. (b) Output waveform of the tested SD APD detector shows a single avalanche rising above the capacitive response residual. The red dashed line represents the discrimination level, which is set to be \SI{25}{\milli \volt}.}
			\label{1}
		\end{figure}
		
		Figure~\ref{1}(a) shows the schematic circuit of the tested SD APD detector. A DC bias voltage combined with the periodic gating signals is reversely loaded on the APD. When the reversed bias voltage is higher than the breakdown voltage, the APD works on Geiger mode where a single photon can result in detectable macroscopic avalanche current. However, the repetition rate of the gating signal is so fast that weak avalanche signals are often buried within the APD's capacitive response~\cite{yuan2007}. In order to remove the capacitive response, the SD technique is applied. That is, first divide the response of APD into two halves, then shift one of them by one gate period, and recombine the two halves to cancel the strong capacitive response. The weak avalanche signal processed after the SD technique is shown in Fig.~\ref{1}(b). Through SD technique, only weak avalanche signals and capacitive response residual remain, which can be distinguished by setting a discrimination level.

		\begin{figure}[htbp]
			\centering
			\includegraphics[width=\linewidth]{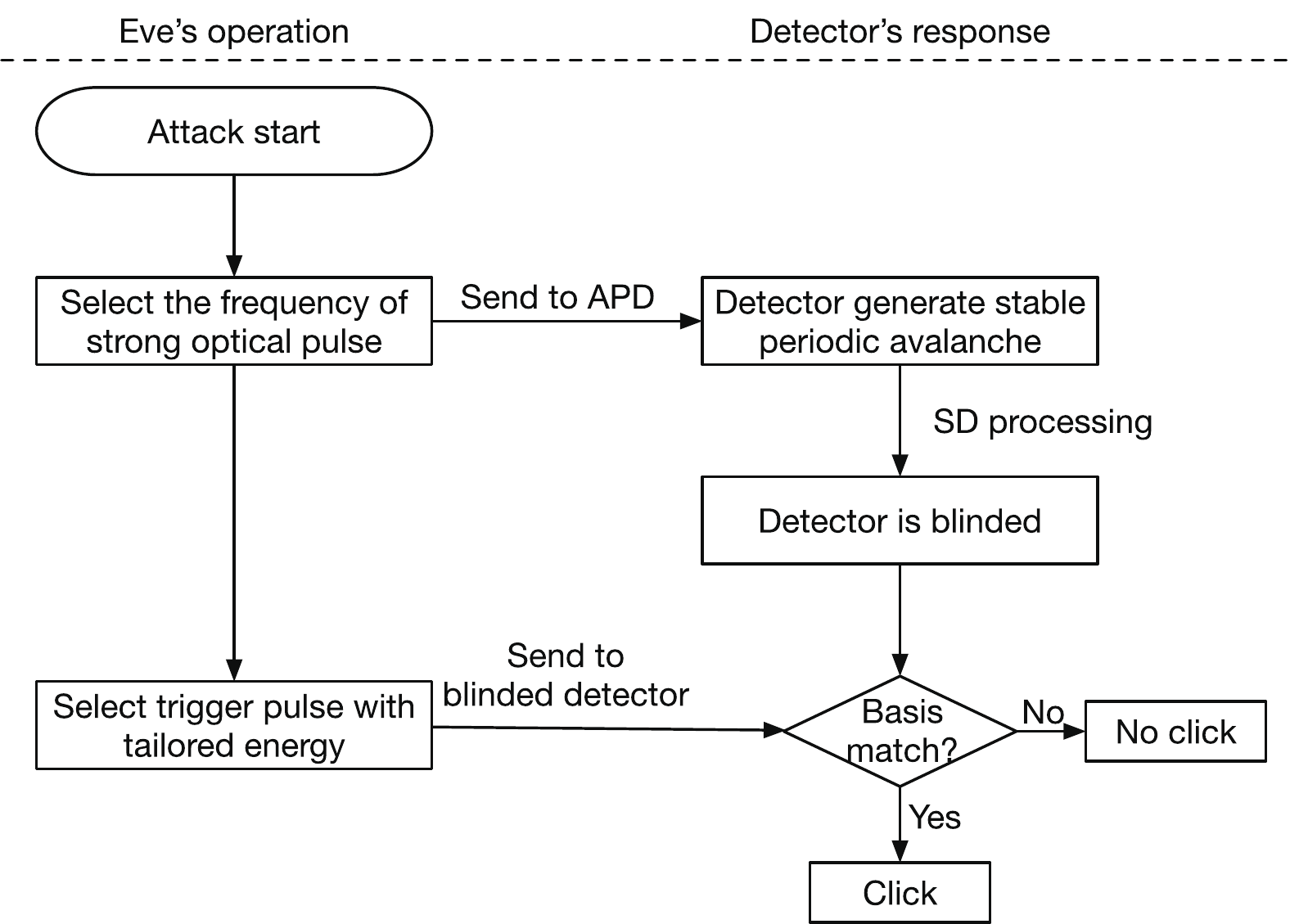}%
			\caption{General process of strong pulse illumination attack.}
			\label{2}
		\end{figure}

		Due to the intrinsic imperfection of SD APD detectors, under the strong c.w.\ light illumination, the SD APD detector might be blinded~\cite{koehlersidki2018}. To eliminate the threat of c.w.\ blinding attack, Ref.~\cite{koehlersidki2018} investigated the behavior of a SD APD under c.w.\ bright-light illumination and proposed practice criteria for practical security of SD APD detectors employed in a QKD system. Under the proposed practice criteria, once Eve uses c.w.\ bright-light to blind SD APD detectors, the large blinding photocurrent exposes the existence of Eve. In addition, the increase of error rate caused by residual capacitive background can also help Bob discover Eve. Therefore, SD APD detectors under this practice criteria can defend c.w.\ bright-light illumination.
		
		However, the effectiveness of this practical criteria under pulsed illumination attack is not fully investigated yet. In this work, we thoroughly test the behavior of SD APD detector under strong pulse illumination. Figure~\ref{2} shows the general process of strong pulse illumination attack. By using the strong optical pulses with the same repetition rate of the gated signal applied to SD APD detector, each optical pulse triggers a stable avalanche photocurrent. The stable and periodic avalanche photocurrent is cancelled out after the SD processing. Therefore, the remaining avalanche photocurrent is lower than the discrimination threshold. As a result, the SD APD detector is blinded and its output detection can be controlled by Eve's classical trigger pulses with tailored energy, which triggers a click only when Bob selects the same basis as Eve. Furthermore, the drop of bias voltage under the strong pulses illumination attack is not as much as that under c.w.\ bright-light attack, which keeps the capacitive response residual being lower than the discrimination threshold. 
		
	\section{EXPERIMENTAL SETUP} \label{Ⅲ}

\begin{figure}[htbp]
			\centering
			\includegraphics[width=\linewidth]{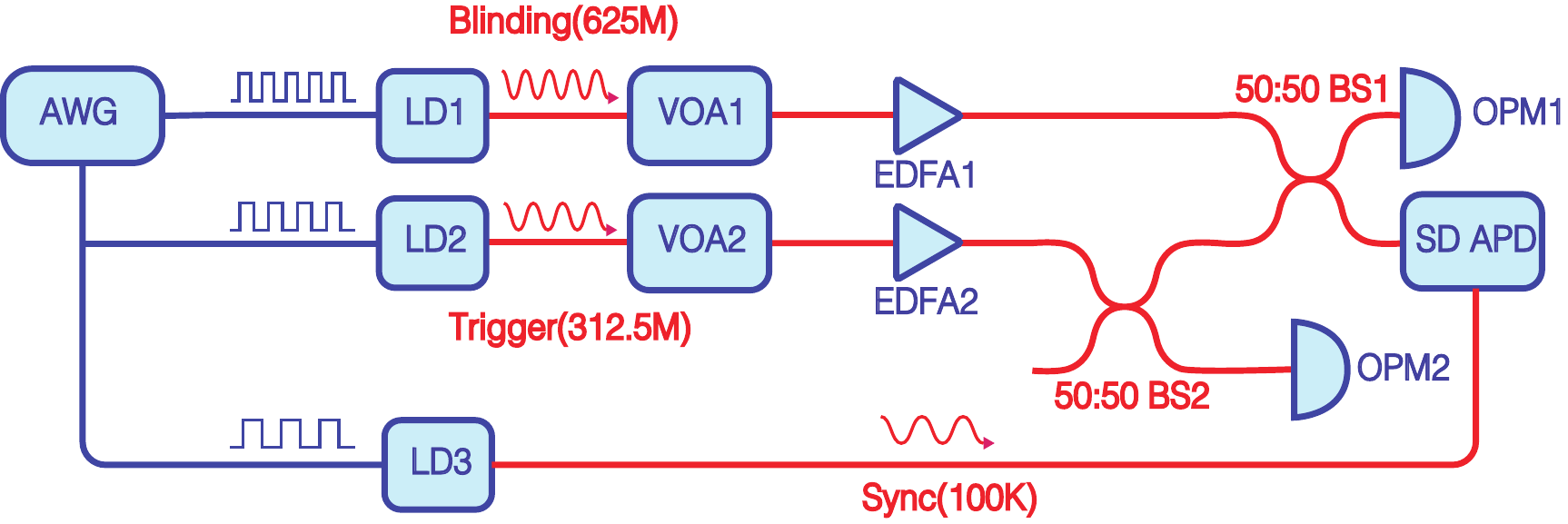}%
			\caption{Schematic diagram of experimental setup. The red lines represent the optical signal, and the blue lines represent the electrical signal. AWG, arbitrary wave generator; LD, laser diode; VOA, variable optical attenuator; EDFA, Erbium-doped fiber amplifier; BS, beam splitter; OPM, optical power meter; SD APD, self-differencing avalanche photodiode detector. As the testing target, the SD APD connected in series with a \SI{1}{\kilo \ohm} bias resistor works at gating frequency of \SI{625}{\MHz}.}
			\label{3}
		\end{figure}

		In order to experimentally explore the behavior of SD APD detectors under strong pulse illumination, the test is conducted using the setup shown in Fig.~\ref{3}. An arbitrary wave generator~(AWG) is used to drive laser diodes. The laser diode 1 (LD1) is driven to emit blinding pulses at \SI{1550}{\nano \meter}, whose repetition frequency is \SI{625}{\MHz} as the same as that of the gating signal applied to the SD APD detector under test. Similarly, the laser diode 2~(LD2) is driven by the AWG to generate \SI{312.5}{}-\SI{}{\MHz} trigger pulses used to control the blinded SD APD detector. The laser diode 3~(LD3) emits pulses with repetition frequency of \SI{100}{\kilo \hertz} to synchronize the attacking setup and the SD APD detector under test to ensure that blinding pulses and trigger pulses can stably illuminate inside the gating period of the SD APD. Variable optical attenuators~(VOAs) and erbium-doped fiber amplifiers~(EDFAs) are used to tune the optical intensity of blinding pulses and trigger pulses. The optical power meter 1~(OPM1) monitors the optical power of the blinding pulses and the optical power meter 2~(OPM2) serves to monitor the optical power of the trigger pulses. Meanwhile, the 50:50 beam splitter 1~(BS1) merges the blinding pulses and the trigger pulses. 

		The SD APD detector under tested is cooled down to \SI{-40}{\degreeCelsius} and applied by \SI{64.2}{\volt} bias voltage. As shown in Fig.~\ref{1}(a), the gating frequency of the APD is \SI{625}{\MHz} and the bias resistor connected in series is \SI{1}{\kilo \ohm}. The resistance value of the bias resistor satisfies the requirement b) of the practice criteria in Ref.~\cite{koehlersidki2018}, which recommends to avoid using a bias resistor exceeding \SI{50}{\kilo \ohm}. It is important to note that we realize the SD operation of avalanche signals by means of software processing instead of practical SD circuit. Compared to the physical realization, software processing can remove the effect of the timing jitter, which makes the result of the SD more precise. 
		
\begin{figure}[htbp]
			\centering
			\includegraphics[width=\linewidth]{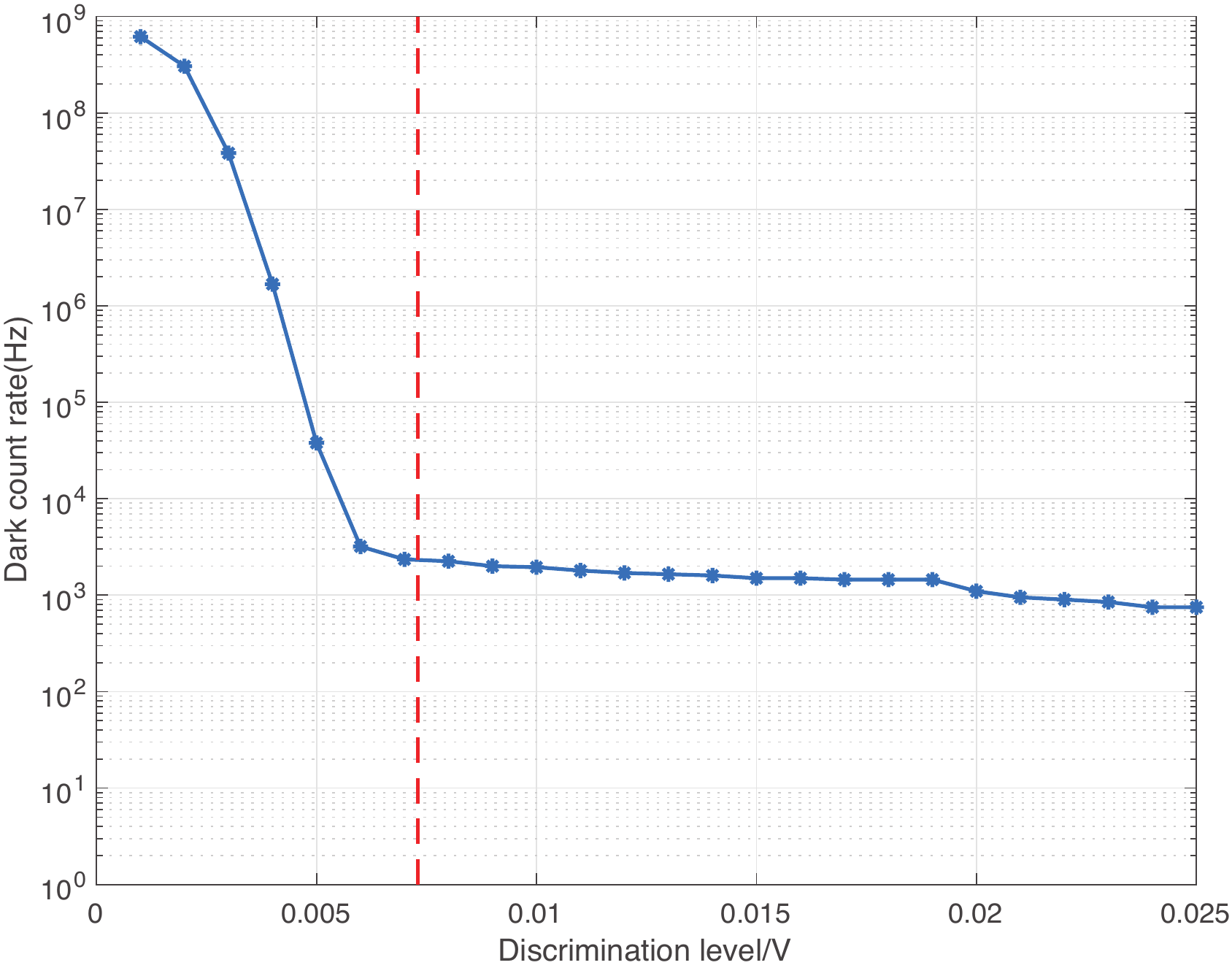}%
			\caption{Dark count rate as a function of the discrimination level. When the discrimination level is lower than \SI{6}{\milli \volt}, the dark count rate mainly comes from the capacitive response residual. Otherwise, the dark avalanches are the major source of the dark count rate. The red dashed line represents the minimum average value of peak amplitude when the SD APD detector is blinded.}
			\label{4} 	
		\end{figure}

		Setting an appropriate discrimination level can not only improve the detection efficiency of the SD APD detector, but also perceive the reduction of excess voltage~\cite{koehlersidki2018}. Therefore, the choice of discrimination level of a SD APD detector is important. Figure.~\ref{4} shows the dark count rate as a function of the discrimination level. As observed from Fig.~\ref{4}, there is a kink at discrimination level of \SI{6}{\milli \volt}, indicating the dark avalanches replace the capacitive response residual to be the dominant contribution to the measured dark count rate when the discrimination level is higher than \SI{6}{\milli \volt}. Therefore, weak avalanche signal and capacitive response residual can be distinguished when the discrimination voltage is higher than \SI{6}{\milli \volt}. However, for a SD APD detector working in a real QKD system, the change of working environment may introduce extra electronic noise, and the detection noise of the SD APD may increase during long-time running. Thus, the set discrimination level not only needs to distinguish between capacitive response residual and weak avalanche signal, but also needs to resist the noise caused by above reasons. To enhance noise resistance of the SD APD detector, the discrimination level is set to \SI{25}{\milli \volt} by the third party who provides the SD APD detector.
		
	\section{EXPERIMENT RESULTS} \label{Ⅳ}
		
		In this study, we conduct an attacking experiment on Bob's SD APD detector with strong optical pulse. LD3 is firstly turned on to send \SI{100}{\kilo \hertz} synchronizing pulses to the SD APD detector for synchronizing the whole testing setup. Then, LD1 is switched on to generate \SI{625}{}-\SI{}{\MHz} blinding pulses, whose intensity is modulated by VOA1 and EDFA1, to illuminate the SD APD detector. Under each intensity of the incident pulses, we measure the avalanche signal after SD processing, collect 480 consecutive periods, and make statistics on the peak amplitude in each period. 
		
\begin{figure}[htbp]
			\centering
			\includegraphics[width=\linewidth]{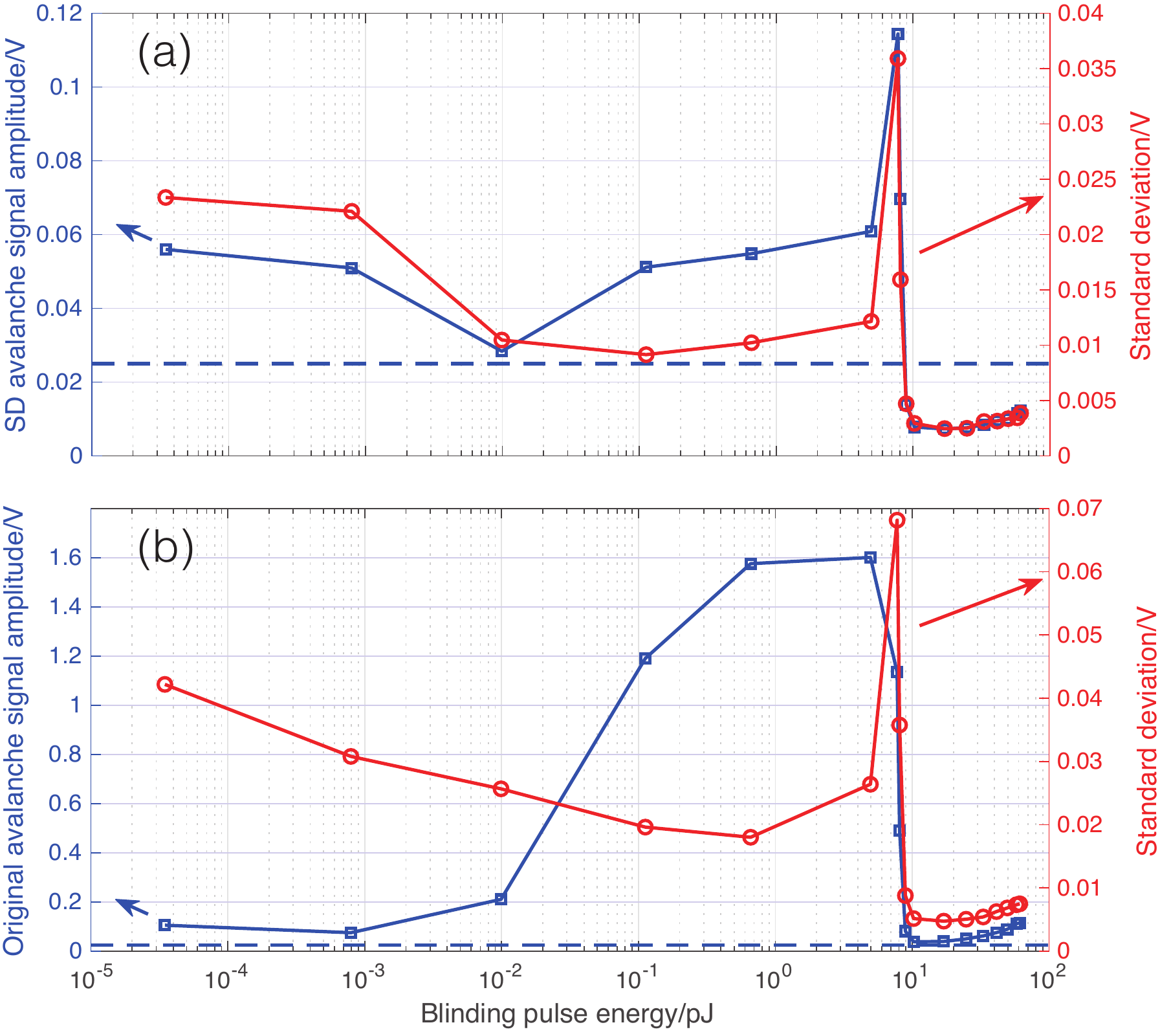}
			\caption{Average value and standard deviation (a) of SD avalanche signal's peak amplitude and (b) of original avalanche signal's peak amplitude as a function of the blinding pulse energy. The blue dashed line represents the discrimination level. When the peak amplitude of SD avalanche signal is lower than it, the detector is blinded.}
			\label{5} 	
		\end{figure}

		Figure~\ref{5}(a) shows the average value of peak amplitude $\bar V_\text{peak}^\text{SD}$ and standard deviation $\sigma^\text{SD}$ of the SD avalanche signal depending on the energy of each blinding pulse. When the blinding pulse energy is small, the $\bar V_\text{peak}^\text{SD}$ of the SD avalanche signal is higher than the discrimination level, \SI{25}{\milli \volt}. By gradually increasing the blinding pulse energy, the $\bar V_\text{peak}^\text{SD}$ of the SD avalanche signal firstly decreases and then starts increasing at \SI{0.01}{\pico \joule} blinding pulse. Remarkably, there is a deep when the blinding pulse energy is \SI{0.01}{\pico \joule}. It is because for this amount of energy, each pulse almost triggers an avalanche in each period, resulting in a smaller amplitude remaining after SD processing. When the blinding pulse energy is higher than \SI{7.76}{\pico \joule}, the $\bar V_\text{peak}^\text{SD}$ of the SD avalanche signal begins to decrease rapidly. Finally, the $\bar V_\text{peak}^\text{SD}$ of the SD avalanche signal is lower than the discrimination level when the blinding pulse energy is higher than \SI{8.92}{\pico \joule}. It means that the SD APD detector can be directly blinded by lowering the amplitude of SD avalanche signal under the strong pulse illumination. After the SD APD detector is blinded, even though the average power of the blinding pulse is increased to \SI{61.09}{\pico \joule}, the count rate of the SD APD detector still does not recover, indicating that the SD APD detector can be blinded stably. 
		
\begin{figure*}[htbp]
		    \centering
		    \includegraphics[width=\linewidth]{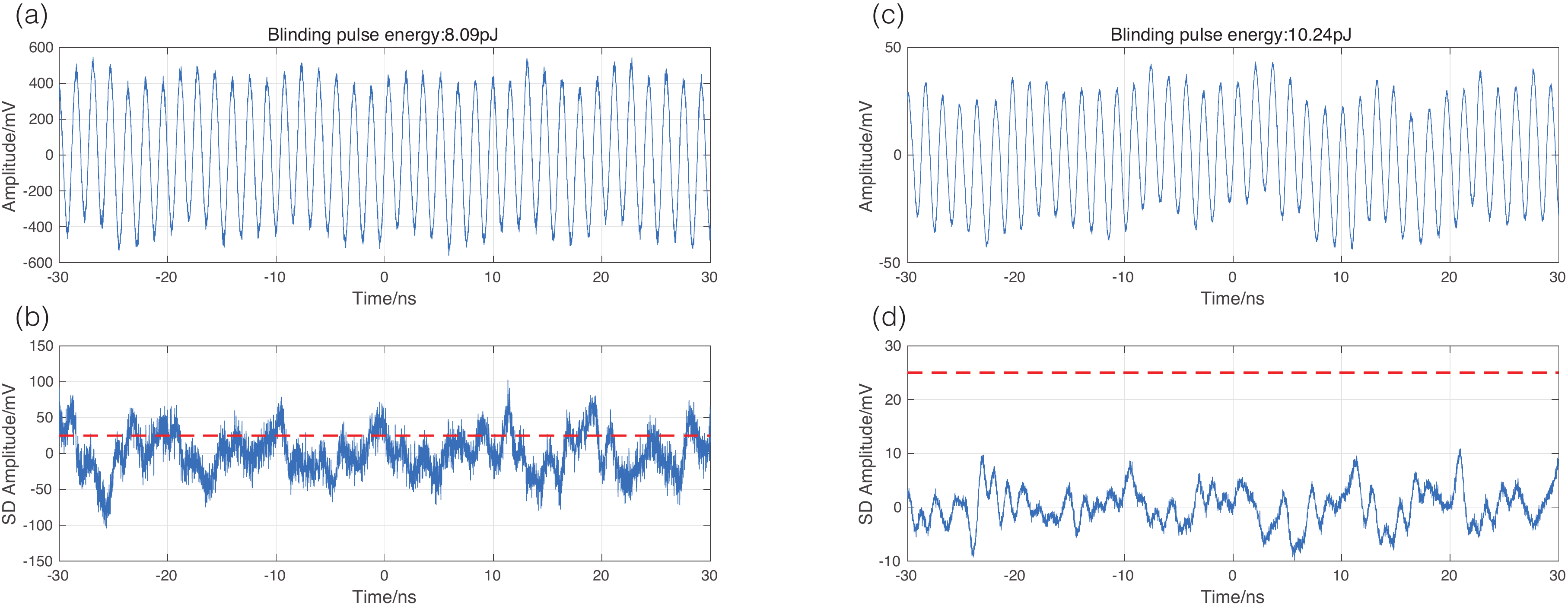}
		    \caption{The amplitude of avalanche signal under specific blinding pulse energies. (a) The waveform of original avalanche signal and (b) the waveform of SD avalanche signal when the blinding pulse energy is \SI{8.09}{\pico \joule}, in which case the SD APD detector is not blinded. (c) The waveform of original avalanche signal and (d) the waveform of SD avalanche signal when the blinding pulse energy is \SI{10.24}{\pico \joule}, in which case the SD APD detector is blinded. The red dashed line represents the discrimination level.}
		    \label{6} 
		\end{figure*}

	    To further understand the blindness of SD APD detector. We conduct the same statistics on the original avalanche signal before SD processing. Figure~\ref{5}(b) shows the $\bar V_\text{peak}$ and $\sigma$ of original avalanche signal as a function of each blinding pulse's energy. With the increasing energy of blinding pulse, $\bar V_\text{peak}$ of the original avalanche signal first increases and then rapidly decreases to \SI{81}{\milli \volt} at \SI{8.92}{\pico \joule}, in which case the SD APD detector is blinded. Figure~\ref{6}(a) and (c) respectively show in detail the amplitude of original avalanche signal when blinding pulse energy is \SI{8.09}{\pico \joule} and \SI{10.24}{\pico \joule}, which are the cases right before and after blinding happened. In Fig.~\ref{6}(a), amplitude is relatively large and the waveform is very unstable, which results that the SD amplitude is higher than the discrimination level. With the increase of energy, the amplitude of the original avalanche signal in each period become smaller and is very stable in Fig.~\ref{6}(c). It means strong pulse illumination lowers amplitude and fluctuation the original avalanche signal, consequently, blinding the SD APD detector.
	   
\begin{figure}[htbp]
			\centering
			\includegraphics[width=\linewidth]{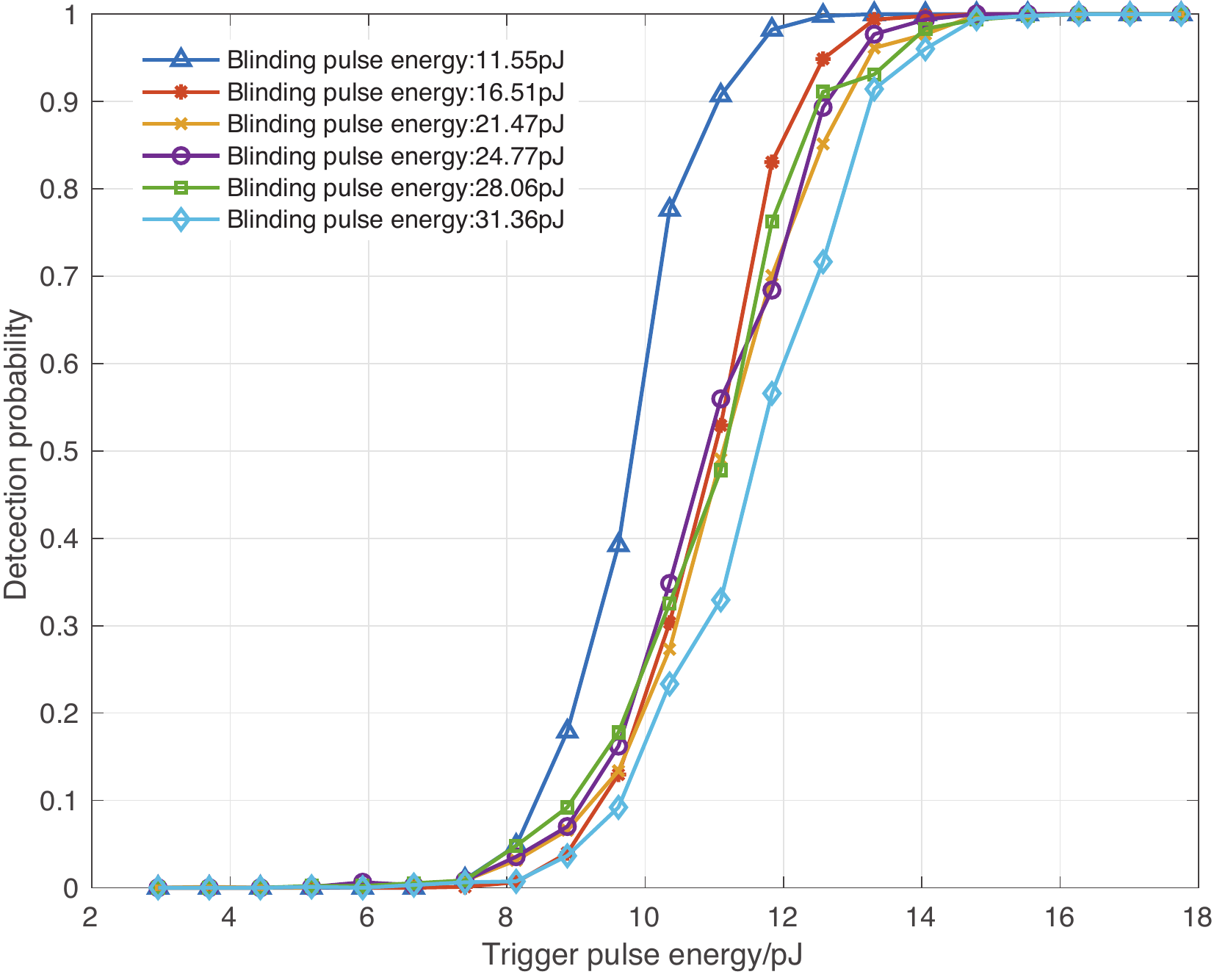}
			\caption{The detection probability as the function of trigger pulse energy under different amounts of blinding pulse energy.}
			\label{7} 	
		\end{figure}

		After the SD APD detector is blinded, in order to control detection outcome of the SD APD detector, LD2 is turned on to send \SI{312.5}{}-\SI{}{\MHz} trigger pulses to the SD APD detector. The trigger pulses are superimposed on blinding pulses through BS1. For each trigger pulse energy, LD2 sends trigger pulses of 960 periods in total. The number of SD avalanche signal's amplitude exceeding the discrimination level is immediately afterwards counted, so as to obtain the detection probability. Figure~\ref{7} shows the detection probability as the function of trigger pulse energy under different amounts of blinding pulse energy, which indicates that the detection probability can vary from 0\% to 100\% with the increase of trigger pulse energy. 
		
		Therefore, Eve can obtain the key by conducting a fake-state attack~\cite{lydersen2010a}. Specifically, Eve firstly intercepts the single photon sent by Alice and randomly selects a basis to measures it as Bob does. Then she resends Bob a trigger pulse superimposed on the blinding pulse according to the measurement result. If Bob's basis choice is consistent with Eve, there could be a detection event. Otherwise, no SD APD detector clicks. Finally, Eve can acquire the identical final key by monitoring the classical channel between Alice and Bob.
		
		Although the trigger pulse enables the detection probability to reach 100\% in the case of high energy, in order to conduct a perfect eavesdropping, the energy of the trigger pulse needs to satisfy the requirement in a BB84 QKD system proposed in Ref.~\cite{lydersen2010a}, which can be expressed as
			\begin{equation}
				E_{always} \leq 2 \times E_{never}, \label{a}
			\end{equation}
		where $E_{always}$ and $E_{never}$ represent the energy of a trigger pulse when the detection probability is 100\% and 0\%, respectively. In the experiment, the case that blinding pulse energy is \SI{11.55}{\pico \joule} satisfies this requirement. When the trigger pulse energy is lower than \SI{6.656}{\pico \joule} the detection probability is 0\%. At the same blinding pulse energy, the maximum energy of trigger pulse that Eve can send is \SI{13.312}{\pico \joule}, and the corresponding detection probability of SD APD detector is 100\%. Therefore, Eve can completely control the output of SD and does not increase the error rate of the final key, which does not expose the existence of Eve. Similarly, for blinding pulse energy of \SI{16.51}{\pico \joule}, \SI{21.47}{\pico \joule}, \SI{24.77}{\pico \joule}, \SI{28.06}{\pico \joule}, and \SI{31.36}{\pico \joule}, the corresponding maximum detection probability is 99.37\%, 27.3\%, 7.08\%, 9.2\%, and 3.67\% respectively when no QBER is introduced. It is notable that these non-100\% detection probabilities can be hidden by the channel loss during the fake-state attack.

	\section{DISCUSSION} \label{Ⅴ}
	
		So far some investigations have contributed to the security of SD APD detectors. In Ref.~\cite{jiang2013}, researchers disclosed a type of pulsed blinding method, in which $\bar V_\text{peak}^\text{SD}$ is at a relatively large value. Therefore, fluctuation in the blinding pulses may cause avalanche signal amplitude to overcome discrimination level, making detector resume counting again. However, in our experiment, by using blinding pulse with higher energy intensity, we directly lower the $\bar V_\text{peak}^\text{SD}$ of avalanche signal. Compared to the previous blinding attack, pulse illumination attack demonstrated in this work drastically reduces the influence of optical power fluctuation and makes the detector blinding more stable.
	
		Significantly, pulse illumination attack might partially invalidate practice criteria proposed in Ref.~\cite{koehlersidki2018}. First, for the criterion of monitoring the photocurrent~\cite{koehlersidki2018}, although it is an effective method to defend c.w.\ bright-light attack, it may be bypassed by a group of blinding pulses that are used in the pulse illumination attack. Specifically, a group of blinding pulses accumulatively introduces a high photocurrent may be averaged before sensing by a photocurrent monitor~\cite{Wu_2020}. Thus, the instant high photocurrent may lower the bais voltage across the APD to blind the SD APD detector. Secondly, for the criterion of avoiding use a quenching or biasing resistor with resistance value higher than \SI{50}{\kilo \ohm}~\cite{koehlersidki2018}, even though the bias resistance of the tested SD APD detetor is only \SI{1}{\kilo \ohm} that satisfied the requirement, strong pulse illumination still blind the SD APD detector. 
		
\begin{figure}[htbp]
			\centering
			\includegraphics[width=\linewidth]{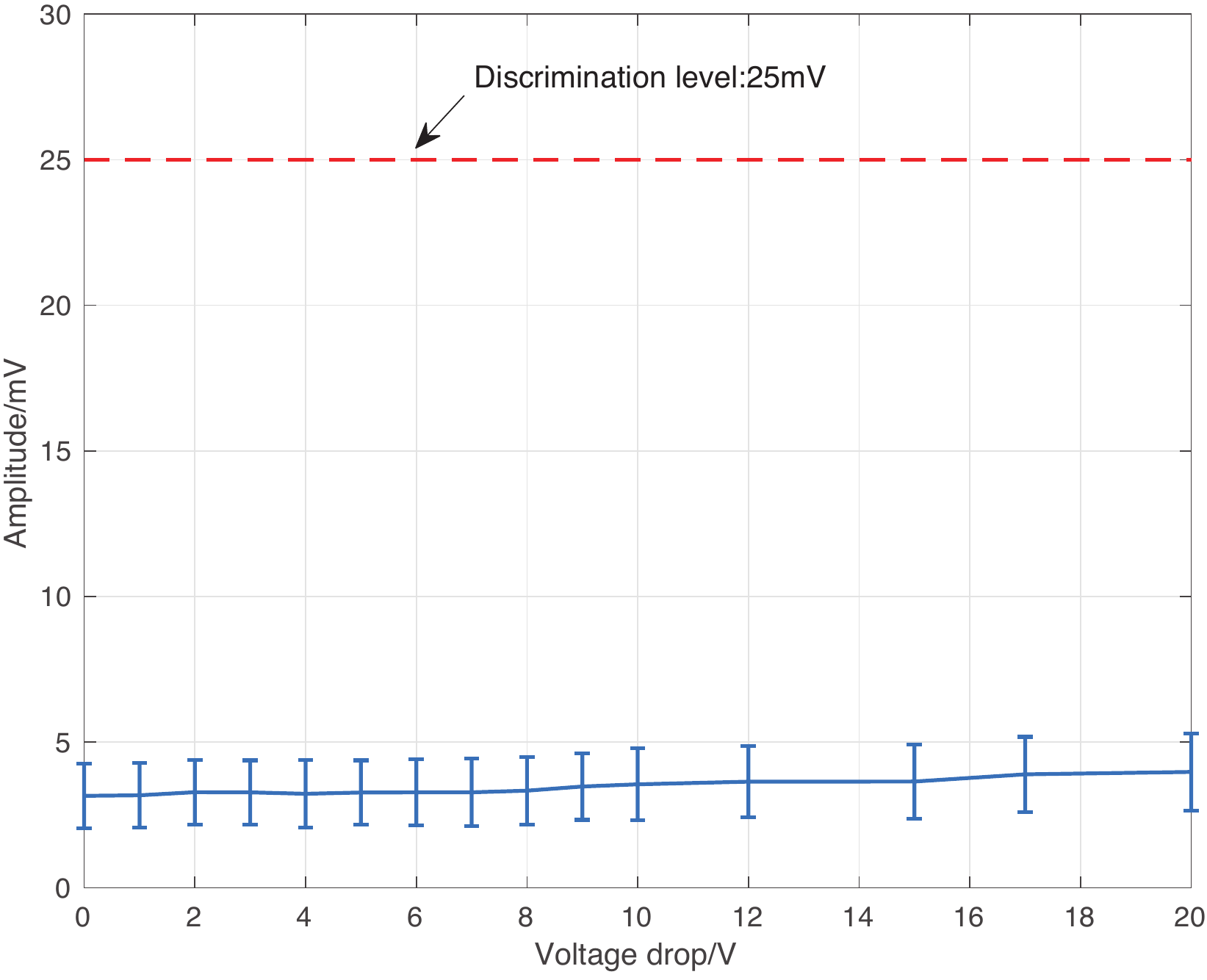}%
			\caption{APD capacitive response measured before the SD processing as a function of the DC bias reduction below its normal value. The red dashed line represents discrimination level, and the blue line represents average amplitude and standard deviation of capacitive response.}
			\label{8} 	
		\end{figure}

		Thirdly, according the requirements c) and e) proposed in Ref.~\cite{koehlersidki2018}, setting a well-selected discrimination level can perceive the reduction of excess voltage through the residual capacitive background, because the capacitive response residual can overcome the discrimination level when the APD's reverse bias voltage decreases~\cite{koehlersidki2018}. However, for the tested SD APD detector, the capacitive response residual does not greatly increase to overcome the discrimination level with the reduction of excess voltage. We perform an experiment to explore the relationship between voltage drop and the SD APD capacitive response measured before SD processing. By varying the DC reverse bias voltage of the APD, the capacitive response of the APD is measured under dark condition. For each set voltage drop, the capacitive amplitude of 1920 consecutive periods is recorded, and we make statistics on the peak values of the capacitive amplitude in each period. As shown in Fig.~\ref{8}, by decreasing the bias voltage of SD APD, the amplitude of APD capacitive response does not increase greatly and is far below the discrimination level. It is different from the explanation proposed in Ref.~\cite{koehlersidki2018}. To understand the origin of the discrepancy, we analyze the internal circuit of the SD APD detector. The failure to recover the count rate may be due to the existence of a filter in the circuit. The capacitive response is filtered in advance, thus reducing the influence of the capacitive response residual. Although the filter can help better distinguish weak avalanche signal from the capacitive response, it also leads to the SD APD being blinded under strong pulse illumination.
		
		Based on the experimental testing, we propose a list of criteria as follows to resist the pulse illumination attack on SD APD detectors.
		\begin{enumerate}
		\item[a)] Remove bias resistor. Under strong pulse illumination, reverse bias voltage across the APD is reduced rapidly when there a is bias resistor in the circuit. 
		\item[b)] Use an optical power limiter or optical fuse. Adding a special passive component, such as optical power limiter~\cite{Zhang2021} or optical fuse~\cite{Todoroki2004}, to sense and response instant high-optical power at the SD APD detector's input can prevent strong pulse from passing through.
	    \item[c)] Remove possible filtering component. The filtering device existing in the circuit makes capacitive response residual ignore the the reduction of excess voltage of APD, which leads to SD APD detectors being blinded stably.
	    \item[d)] Set an appropriate discrimination level. No matter temperature change of working environment or long time running, doing so not only ensures that the SD APD is not disturbed by noise, but also enables capacitive response residual overcomes the discrimination level when excess voltage reduces.
		\end{enumerate}

	\section{CONCLUSION} \label{VI}
		In summary, we experimentally investigate the behavior of SD APD under strong pulse illumination. We show that strong pulse illumination can hack SD APD detectors in high-speed quantum key distribution systems to learn the secret key without introducing extra QBER. Based on the testing results, we find that strong pulse illumination attack presents blinding stability and strong optical pulse can be used as a new tool for eavesdropper to blind the SD APD detector. Meanwhile, we propose a list of criteria to enhance the practical security of SD APD detectors. This work greatly contributes to improve the security of the practical high-speed QKD system.

	\begin{acknowledgments}
	We thank Konstantin Zaitsev and Vadim Makarov for helpful discussions. This work was funded by the National Natural Science Foundation of China (grants 61901483 and 62061136011), the National Key Research and Development Program of China (grant 2019QY0702), and the Research Fund Program of State Key Laboratory of High Performance Computing (grant 202001-02).
	\end{acknowledgments}
	
	\def\bibsection{~}
	\bibliography{article}

\end{document}